# On the nature of screening in charge regulated macroion solutions

Sunita Kumari
*Department of Physics, Indian Institute of Technology, Jodhpur 342037, India*

Rudolf Podgornik*
*School of Physical Sciences and Kavli Institute for Theoretical Sciences,
University of Chinese Academy of Sciences, Beijing 100049, China
Wenzhou Institute, University of Chinese Academy of Sciences, Wenzhou, Zhejiang 325000, China and
CAS Key Laboratory of Soft Matter Physics, Institute of Physics,
Chinese Academy of Sciences, Beijing 100190, China†*

We present a derivation of the screening length for a solution containing a charge-regulated macroion, *e.g* protein, with its counterions. We show that it can be obtained directly from the second derivatives of the total free energy by taking recourse to the "*uncertainty relation*" of the Legendre transform which connects the Hessians or the local curvatures of the free energy as a function of density and its Legendre transform, i.e., osmotic pressure, as a function of chemical potentials. Based on the Fowler-Guggenheim-Frumkin model of charge regulation we then analyze the "screening resonance" and the "overscreening" of the screening properties of the charge-regulated macroion solution.

## I. INTRODUCTION

Electrostatic interactions are a fundamental component of the Deryagin-Landau-Verwey-Overbeek (DLVO) paradigm of molecular forces in the colloid [1] and nanoscale domains [2], being of particular importance for various phenomena in the biological and biomolecular context [3]. The mean-field formulation of the Poisson-Boltzmann (PB) theory is standardly used for describing the nanoscale electrostatics [4]. While the PB paradigm has well-defined limits of applicability, stemming either from the model [5] and/or from the methodology [6, 7], it is still used as a preferred estimation tool, often providing a quantitatively correct form of the electrostatic interaction [8].

In many contexts, but most importantly for protein and complex macroion solutions, the usual PB paradigm assumption of constant surface charge density or constant surface potential [9] breaks down, as the dissociable molecular moieties respond to their environment, especially to the presence of each other, in a way that modifies both the charge density as well as the surface potential, adjusting them according to the separation between them and the bathing solution conditions [10]. This conceptual framework is formally referred to as the *charge regulation* (CR) [11] and is known to play a fundamental role in, *e.g.*, macromolecular interactions [12–16], protein solutions [17–19] and protein interactions [20–22], as well as polyelectrolytes and polyelectrolyte gels [23–28].

Screening of electrostatic interactions is one of the fundamental properties of electrolyte solutions, standardly quantified in terms of the *screening length* [29], that has received renewed attention in recent years connected with the anomalous screening in, *e.g.*, ionic liquids [30, 31]. In the PB framework, the screening length is identified as the *Debye length*, $\lambda_D$, and is obtained from linearizing the PB equation as [11, 32]

$$\lambda_D^{-2} \equiv \kappa_D^2 = (\varepsilon_w \varepsilon_0)^{-1} \frac{\partial \rho(\psi)}{\partial \psi} \bigg|_{\psi=0} \qquad (1)$$

with $\rho(\psi)$ being the mobile charge density as a function of the mean-field electrostatic potential $\psi$, and $\varepsilon_w, \varepsilon_0$ are the dielectric constant of solvent (water) and the dielectric permittivity of vacuum. The $\psi = 0$ indicates a bulk value. The derivative of the charge density w.r.t the potential is, of course, nothing but the capacitance density at thermal equilibrium so that the inverse square of the screening length can then be identified as the capacitance density, *i.e.*, the response function for the charge density in the presence of an imposed electrostatic potential difference [33]. This can be furthermore expressed in terms of the thermal fluctuations of the electrostatic potential by the Einstein formula [29]. For a solution of negatively charged macroions (M) and their univalent counterions (+), one obtains

$$\kappa_D{}^2 \equiv \frac{4\pi\ell_B}{e^2}\Big(e^2 n_+(\psi) + Q_M^2 n_M(\psi)\Big)\Big|_{\psi=0}, \tag{2}$$

where $n_{+,M}(\psi = 0)$ are the concentrations of the counterions (charge $+e$) and macroions (charge $-Q_M$) in the bulk ($\psi = 0$), and $\ell_B$ is the Bjerrum length, equal to the distance at which two unit charges interact with thermal energy $k_B T$ and given by $\ell_B = e_0^2/4\pi\varepsilon_w\varepsilon_0 k_B T$ (in water at room temperature, the value is $\ell_B \approx 0.7$ nm). In the symmetric case, $e_+ = Q_M$ the above formula reduces to the standard Debye expression.

---

* podgornikrudolf@ucas.ac.cn
Also affiliated with Department of Physics, Faculty of Mathematics and Physics, University of Ljubljana, 1000 Ljubljana, Slovenia
† podgornikrudolf@ucas.ac.cn

We recently showed [29] that the screening length is fundamentally modified in ionic solutions which cannot be described within the PB paradigm but require more complicated Voorn-Overbeek type free energy functionals applicable to macroion solutions. In that case, the screening length was found to exhibit a non-monotonic under-screening behavior as a function of the size of the macroion, displaying a different functional dependence on the volume fractions of the components than the standard Debye screening length [29].

Motivated by these findings we now examine the nature of ionic screening in solutions of *charge regulating* (CR) macroions. In fact, Avni *et al.* [11] derived a formula for the screening length in an ionic solution composed of simple salt ions and charge-regulated macroions, which can be cast for the case of a macroion and its counterion as

$$\lambda_D{}^{-2} \equiv \frac{4\pi\ell_B}{e^2}\Big(e^2 n_+(\psi) + Q_M(\psi)^2 n_M(\psi) + n_M(\psi)(\Delta Q_M(\psi))^2\Big)\Big|_{\psi=0}, \tag{3}$$

where in addition to $n_{+,M}(\psi = 0)$, we also have the bulk concentration of the CR macroions, $n_M(\psi = 0)$, the bulk value of the macroion charge, $Q_M(\psi = 0)$, as well as the fluctuations around this average due to charge regulation mechanism

$$(\Delta Q_M(\psi = 0))^2 \equiv -k_B T \frac{\partial Q_M(\psi)}{\partial \psi}\Big|_{\psi=0}. \tag{4}$$

A variant of this expression was already stated in the work of Jönsson and Lund [17, 34]. Of the four terms in the above equation, it is the last one that most obviously defies the Debye form of the screening length. Eq. 3 also suggests that the screening properties of the system can be formulated in terms of the total capacitance: the bulk capacitance, stemming from the spatial redistribution of all the charged particles, including the CR macroions, ($n_{+,-}$ and $n_M$), and the intrinsic CR capacitance due to the charge regulation mechanism proportional again to $n_M$. Since the simple salt ions have a fixed charge, they contribute only to the bulk capacitance, whereas the macro-ions have additive contributions to both, the bulk as well as the intrinsic capacitance.

In what follows we will apply the properties of Legendre transforms to translate between the standard formulation in terms of the electrostatic potential, and a dual formulation in terms of the electric displacement field [35, 36] in order to derive an alternative form of the screening length. This form of the screening length based on the properties of the Hessian of the free energy and its Legendre transform will allow us to bypass the PB equation and straightforwardly obtain the contribution of different CR mechanisms to the screening properties of CR macroion solutions. We will specifically analyze the screening length as a function of the solution composition and the parameters of the CR models. We will identify some salient effects of the CR on screening properties.

## II. LEGENDRE TRANSFORM FORMULATION OF MEAN-FIELD CHARGE REGULATION THEORY

We first specify a simple model of a CR macroion solution, a modification of those previously described in the literature [32, 37]. The system under consideration is assumed to be composed of three components: the CR macroions of volume fraction $c_1$ with $N_1$ negatively charged sites, free positively charged counterions of volume fraction $c_2$, and positively charged counterions residing on some of the $N_1$ sites of the macroion surface with an annealed dissociation fraction $0 \leq \phi \leq 1$ [32]. We furthermore assume that the counterions in the solution and counterions on the macroion surface are characterized by the same chemical potential as they correspond to the same counterion species that can be either dissolved in the solution or residing on the macroion surface. The

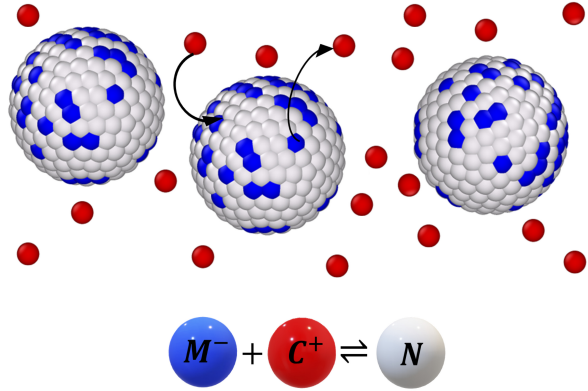

Figure 1: A schematic representation of the charge-regulated solution of charge-regulated macroions and counterions, showing three macroions and their counterions. Charge regulated macroions of concentration $c_1$, with fixed negative charges (blue sites), onto which positively charged counterions of concentration $c_2$ (red sites) can adsorb, neutralizing them (gray sites), corresponding to the reaction ● + ● ⇋ ●. The fraction of the neutralized sites of the macroion is $\phi$, so that the effective charge of the macroion is $-eN_1(1-\phi)$, varying from 0 to $-eN_1$ as the fraction $\phi$ varies from 1 to 0.

*free energy density* in thermal units can then be written as a sum of three terms

$$f(c_1, c_2, \phi) = c_1 \left(\log(c_1 a^3) - 1\right) + c_2 \left(\log(c_2 a^3) - 1\right) + c_1 \, g(\phi), \tag{5}$$

where the first term corresponds to macroions, the second to solution counterions, and the last one to macroion charge regulation. $g(\phi)$ is the free energy describing the CR process, which can be either the dissociation of the macroion surface groups or the adsorption of solution ions onto the macroion surface. For simplicity, we also assume that the macroion and the counterion are of the same size, $a$, an assumption that is easy to relax but leads to an additional parameter that we want to avoid at this point.

We next assume that the CR process can be described by the Fowler-Guggenheim-Frumkin isotherm [38–40] which implies

$$g(\phi) = N_1 \tilde{g}(\phi) = N_1 \left(-\phi\alpha + (\phi \log \phi + (1-\phi)\log(1-\phi)) + \tfrac{1}{2}\chi\phi^2\right), \tag{6}$$

where $N_1$ is the number of dissociable sites on the macroion surface. Without the last term, the above free energy describes the adsorption with an energy cost $\alpha$ and entropy given by the lattice gas model. $\alpha \geq 0$ ($\alpha \leq 0$) corresponds to an attraction (repulsion) between counterions and the macroion. This part implies nothing but the Langmuir adsorption isotherm, while the last term specifically pertains to the pair interaction of adsorbed surface charges and is of the general Flory-Huggins type as used in the regular solution theory or the monomer interactions in polymer solutions [41]. $\chi \geq 0$ ($\chi \leq 0$) corresponds to an effective attraction (repulsion) between the adsorbed counterions.

The general structure of the free energy for the system under consideration, Eq. 5, is then

$$\mathcal{F}(c_1, c_2, \phi) = \int_V d^3\mathbf{r} \, f(c_1, c_2, \phi) = \int_V d^3\mathbf{r} \left(f(c_1, c_2) + c_1 N_1 \, \tilde{g}(\phi)\right), \tag{7}$$

consistent with Markovich *et al.* [32], where $g(\phi)$ is given by Eq. 6. We will now follow closely the formulation of the CR problem for macroscopic surfaces in terms of Legendre transform [36] but applied to a solution of (point) macroions [42]. The thermodynamic potential, including the electrostatic energy as a functional of the dielectric displacement vector $\mathbf{D}$, is then given by [36]

$$\begin{aligned}\mathcal{F}[c_1, c_2, \phi, \mathbf{D}] =& \int_V d^3\mathbf{r} \left(f(c_1, c_2) + c_1 N_1 \, \tilde{g}(\phi) - \mu_1 c_1 - \mu_2 c_2 - \mu_2' c_1 (N_1 \phi)\right) + \\ &+ \int_V d^3\mathbf{r} \left(\frac{\mathbf{D}^2}{2\varepsilon} - \psi \left(\boldsymbol{\nabla} \cdot \mathbf{D} + eN_1 c_1 - ec_2 - ec_1 (N_1 \phi)\right)\right),\end{aligned} \tag{8}$$

where $\varepsilon = \epsilon\epsilon_0$, with $\epsilon$ the dielectric constant of the aqueous solution. Here, we assumed that the chemical potentials $\mu_2$ and $\mu_2'$ are unrelated, but as $c_2$ and $c_2'$ refer to the same ions, either free or residing on the macroion surface, we will eventually identify $\mu_2 = \mu_2'$. In the above free energy, the Poisson equation is implemented with a Lagrange multiplier, being the electrostatic potential $\psi$. The chemical potentials of the species 1, 2 and $2'$ are obtained as

$$\mu_1 = \frac{\partial f}{\partial c_1} + N_1\left(\tilde{g}(\phi) - \mu_2'\phi\right) \qquad \mu_2 = \frac{\partial f}{\partial c_2} \qquad \mu_2' = \frac{\partial \tilde{g}}{\partial \phi}. \tag{9}$$

The effect of electrostatics is then simply to displace all the chemical potentials to electrochemical potentials $\mu_1 \longrightarrow \mu_1 + eN_1\psi$, and $\mu_2 \longrightarrow \mu_2 - e\psi$ and $\mu_2' \longrightarrow \mu_2' - e\psi$. The thermodynamic equilibrium state corresponds to the minimum of the functional $\mathcal{F}[c_1, c_2, \phi, \mathbf{D}]$. Minimizing first with respect to the vector $\mathbf{D}$ we obtain simply the definition of the dielectric displacement vector as $\mathbf{D} = -\varepsilon\boldsymbol{\nabla}\psi$, and minimizing with respect to $\psi$ we get back the Poisson equation

$$\boldsymbol{\nabla} \cdot \mathbf{D} = \rho(\psi) = e_1 c_1 + e_2 c_2 = -eN_1(1-\phi)c_1 + ec_2, \tag{10}$$

where $e_1$ is the charge of the macroion, and $e_2$ is the charge of the counterions. The free energy can now be expressed in terms of the electrostatic and chemical potentials to get the corresponding thermodynamic potential in the form dual to Eq. 8

$$\mathcal{G}[\mu_1, \mu_2, \mu_2', \psi] = -\int_V d^3\mathbf{r} \left(\tfrac{1}{2}\varepsilon(\boldsymbol{\nabla}\psi)^2 + p(\mu_1 - e\psi, \mu_2 + e\psi, \mu_2' - e\psi)\right). \tag{11}$$

where we introduced the Lagrange transform of the free energy, or the osmotic pressure as a function of chemical potentials as

$$-p(\mu_1 - e\psi, \mu_2 + e\psi, \mu_2' - e\psi) = f(c_1, c_2) - (\mu_1 + eN_1\psi + \tilde{p}(\mu_2' - e\psi))c_1 - (\mu_2 + e\psi)c_2, \tag{12}$$

with

$$\tilde{p}(\mu_2' - e\psi) = -N_1\tilde{g}(\phi) + N_1\phi(\mu_2' - e\psi). \tag{13}$$

The CR part pertains to the ions residing on the macroion surface either due to adsorption or dissociation and is thus contained entirely in $\tilde{p}(\mu_2')$. The formal structure of Eq. 12 follows directly from the model-free energy Eq. 5. Its interpretation is as follows: without CR the counterions and macroions are separate species with electrochemical potentials $\mu_1 - e\psi$ and $\mu_2 + e\psi$. When the macroions are charge regulated what happens can be cast as a renormalization of $\mu_1$, to $\mu_1 \longrightarrow \mu_1 - N_1\tilde{g}(\phi) + N_1\phi\mu_2'$, as well as a renormalization of the charge of the macroion, $-eN_1$, to $-eN_1(1-\phi)$. Renormalization of $\mu_1$ simply reflects the fact that the model Eq. 12 adds an additional energy, $N_1\tilde{g}(\phi)$, to the macroion and that the macroion is associated with $N_1\phi$ counterions. Renormalization of the macroion charge stems from the fact that the counterions are charged and those that reside on the surface of the macroion modify its charge. We now proceed to derive the mean-field equations from the functional Eq. 11 and then the ensuing screening length directly from the free energy, bypassing altogether the expansion of the mean-field equations to get the Debye-Hückel limit.

### III. SADDLE-POINT (POISSON-BOLTZMANN) EQUATION AND SCREENING LENGTH

The Legendre transformed free energy Eq. 11 can now be used to derive the saddle-point equation, or equivalently the PB equation. The Euler-Lagrange equation reads

$$\varepsilon\nabla^2\psi = \frac{\partial}{\partial\psi}p(\mu_1 - e\psi, \mu_2 + e\psi, \mu_2' + e\psi)) = (-e)\frac{\partial p}{\partial\mu_1} + (e)\frac{\partial p}{\partial\mu_2} + (e)\frac{\partial p}{\partial\mu_2'} = -\rho(\psi), \tag{14}$$

where $\rho(\psi)$ is the charge density, Eq. 10, since the derivatives on the rhs are evaluated via the Gibbs-Duhem relations.

From the definition of the chemical potentials Eq. 9,

and assuming that the system is in equilibrium with a bulk reservoir so that the chemical potentials are obtained from the volume fractions in the bulk reservoir, we derive the charge density as a function of the bulk volume fractions and the electrostatic potential, giving us the Poisson-Boltzmann equation, Eq. 14. Apart from the different CR model, Eq. 6, with $\chi \neq 0$, the PB equation coincides with the one derived by Markovich *et al* [32]. Generally, it belongs to a PB-type of equations that include charge regulation in the form of a field theory [43].

From the PB equation, Eq. 14, one can derive the screening length by expanding the charge density to the lowest order as a function of the electrostatic potential $\psi$.

$$\kappa^2 \equiv \frac{1}{\varepsilon} \frac{\partial^2}{\partial \psi^2} p\Big(\mu_1 - e\psi, \mu_2 + e\psi, \mu_3 - \tilde{p}(\mu_1' - e\psi, \mu_2 + e\psi)\Big)\bigg|_{\psi=0}. \tag{15}$$

From the definition of the screening length, Eq. 1, as applied to the charge density, Eq. 10, with the Gibbs-Duhem relations, we are then led to the following form of the inverse square of the screening length

$$\kappa^2 = \frac{1}{\varepsilon} \frac{\partial^2 p}{\partial \psi^2} = \frac{1}{\varepsilon} \frac{\partial \rho(\psi)}{\partial \psi}\bigg|_{\psi=0} = \frac{4\pi \ell_B}{e^2} \Big( \underbrace{\sum_{i,j=1,2} e_i e_j \left(\frac{\partial^2 p}{\partial \mu_i \partial \mu_j}\right)}_{Debye\ screening} + \underbrace{c_1\ e^2 \left(\frac{\partial^2 \tilde{p}}{\partial \mu_2'^2}\right)}_{CR\ screenning} \Big), \tag{16}$$

with $e_i$ are defined in Eq. 10. The first, intrinsic part contains all mobile species with their charges assumed to be fixed, thus corresponding to the standard Debye screening, while the second, CR screening part contains only the charges that are exchanged in the charge regulation process. The first derivative is therefore taken at fixed $e_i$ and the second one at fixed $c_2$. In the above formula, we recognize just a different way to write Eq. 3 where the response functions have been written with the Hessian of the osmotic pressure as a function of the respective chemical potentials.

The standard connection between the number fluctuations, $\langle N_i N_k \rangle - \langle N_i \rangle \langle N_k \rangle$, and the pressure derivatives, $\left(\partial^2 p/\partial \mu_i \partial \mu_j\right)$, in the grand canonical ensemble at fixed values of the charges [44] can be applied to the fluctuations of the charge through the fluctuation of the number of particles, $Q_i = e_i N_i$, at fixed $e_i$, or the relationship between $\partial^2 \tilde{p}/\partial \mu_2'^2$ and $\tilde{Q}_2 = eN_1$, at fixed number of sites $N_1$ but fluctuating chartge regulated charge $e$. The square of the inverse Debye length is then composed of a term due to charge fluctuations of cations and macroions, with a fixed number of charges, and another term of the CR macroion. We can then write

$$\kappa^2 = \frac{4\pi \ell_B}{e^2} \Big( \sum_{i,j=1,2} \frac{1}{V} \Big(\langle Q_i Q_k \rangle - \langle Q_i \rangle \langle Q_k \rangle \Big)_{e_i} + c_1 \Big(\langle \tilde{Q}_2^2 \rangle - \langle \tilde{Q}_2 \rangle^2 \Big)_{c_2} \Big), \tag{17}$$

with proportionality factor being $\beta/(\varepsilon V)$. The inverse square of the screening length therefore decouples and is proportional to the sum of the charge fluctuations in solution at fixed charging but fluctuating local density, plus the concentration of macroions times the charge fluctuations due to charge regulation at fixed local density of the macroions. This decomposition is intuitively clear and meaningful: the inverse square of the screening length is just charge fluctuations expressed in different units.

The crucial part of our derivation that will allow us to bypass the PB equation and calculate the screening length directly from the free energy is now based on the "*uncertainty relation*" of the Legendre transform which connects the Hessians or the local curvatures of the free energy as a function of density and its Legendre transform, *i.e.*, the osmotic pressure as a function of the chemical potentials in a manner reminiscent of the Fourier transform uncertainty relation $\Delta x \Delta k \simeq 1$. The exact form of the Legendre transform "uncertainty relation" as written by Zia *et al.* [45] can be straightforwardly generalized to a multicomponent system where it reads [29]

$$\sum_{i,j=1,2} \frac{\partial^2 p}{\partial \mu_i \partial \mu_j} \frac{\partial^2 f}{\partial c_j \partial c_k} = \delta_{ik} \qquad \text{and} \qquad \frac{\partial^2 \tilde{p}}{\partial \mu_2'^2} \frac{\partial^2 \tilde{g}}{\partial \phi^2} = 1, \tag{18}$$

where $p$ is the Legendre transform of $f$, Eq. 12, and $\tilde{p}$ is the Legendre transform of the charge regulation free energy $\tilde{g}$, Eq. 13, that depends on a single variable. The inverse square of the screening length, Eq. 19, then becomes

$$\kappa^2 = \frac{4\pi \ell_B}{e^2} \Big( \sum_{i,j=1,2} e_i e_j \left(\frac{\partial^2 f}{\partial c_i \partial c_j}\right)^{-1} + c_1\ e^2 N_1 \left(\frac{\partial^2 \tilde{g}}{\partial \phi^2}\right)^{-1} \Big), \tag{19}$$

with again $e_1$ is the charge of the macroion, $e_1 = -eN_1(1-\phi)$, and $e_2$ is the charge of the counterions, $e_2 = e$. This is our final formula for the screening length involving only the derivatives of the free energy with respect to concentration variables. So clearly the screening length can be obtained directly from the free energy and its second derivatives. No need to write down the PB equation itself. The above derivation provides an important insight because, with the different CR models, the PB equations become increasingly complex and any simplification in the derivation of the screening length should be welcome.

In the framework of the CR model, given by the free energies Eq. 5 and Eq. 6, the implication of Eq. 19 is then

$$\begin{aligned}\kappa^2 &= \frac{4\pi \ell_B}{e^2}\bigg(\sum_{i,j=1,2} e_i e_j \begin{pmatrix} \frac{1}{c_1} & 0 \\ 0 & \frac{1}{c_2} \end{pmatrix}^{-1} + c_1\, e^2 N_1 \left(\frac{1}{\phi(1-\phi)} + \chi\right)^{-1}\bigg) = \\ &= 4\pi \ell_B \bigg(c_2 + N_1^2(1-\phi)^2 c_1 + c_1 \left(\frac{N_1}{\frac{1}{\phi(1-\phi)} + \chi}\right)\bigg).\end{aligned} \quad (20)$$

This expression can be deconstructed in light of Eq. 16 as follows: the first two terms correspond to the intrinsic part with counterions and macroions with their charges equal to $e$ and $-eN_1(1-\phi)$, respectively, assumed to be fixed at the value given by the CR mechanism for the macroion, while the third, charge regulating part, contains only the charges that are exchanged in the charge regulation process on the macroion. It is straightforward but algebraically tedious to confirm that the above result follows also from the PB equation Taylor expanded to the first order in the electrostatic potential.

## IV. RESULTS AND DISCUSSION

We now study the consequences of the CR model for the dependence of the screening length on the bulk concentration of the macroions. We are specifically interested in the salient features of the screening problem in CR solutions, not in the numerical details.

There are two new phenomena that can occur in strongly CR macroion solutions: *the screening resonance*, an anomalous non-monotonic decrease of the screening length at large negative values o $\chi$, and *the underscreening*, an anomalously long tail of the electrostatic interaction. We use the latter term as it is applied to an electrical double layer of ionic liquids [46].

First, we note that $\mu_2 = \mu_2'$, as the surface cations and the bulk cations are identical, we obtain for the bulk $\psi = 0$ from Eq. 9 the connection $\phi = \phi(c_2)$ with

$$\log c_2 = -\alpha + \log \frac{\phi}{(1-\phi)} + \chi \phi \longrightarrow \phi(c_2) = \frac{1}{1 + e^{-\log(c_2 a^3) - \alpha + \chi\phi}}, \quad (21)$$

with the surface interactions strength, $\alpha$, just renormalizing the value of $c_2$. Finally, the relation between $c_1$ and $c_2$ follows from the electroneutrality condition in the bulk as

$$eN_1(1-\phi) c_1 - ec_2 = 0, \quad (22)$$

yielding the relation $c_1 = c_1(c_2)$, *i.e.*, the volume fraction of the macroions as a function of the volume fraction of the counterions, which we take as the experimentally accessible variable. The dependence of the screening length is thus reduced to $\kappa^2 = \kappa^2(c_2)$, *i.e.*, it depends only on the bulk concentration of the counterions.

The final form of the screening length, or its inverse, is then obtained by using again the electroneutrality condition which provides the connection $c_1(c_2)$ so that Eq. 20 finally yields

$$\kappa^2(c_2)/(4\pi \ell_B) = c_2\Big(1 + N_1(1-\phi(c_2))\Big) + \frac{c_2}{(1-\phi(c_2))}\left(\frac{1}{\frac{1}{\phi(c_2)(1-\phi(c_2))} + \chi}\right), \quad (23)$$

with $\phi = \phi(c_2)$ obtained from the CR model Eq. 21. It would be interesting to compare this with the predicted screening length that does not take into account the charge regulation, *i.e.*, the counterions have a charge $e$, while the macroions are assumed to have the full charge, $-N_1 e$, formally corresponding to the limit $\phi \longrightarrow 0$. In

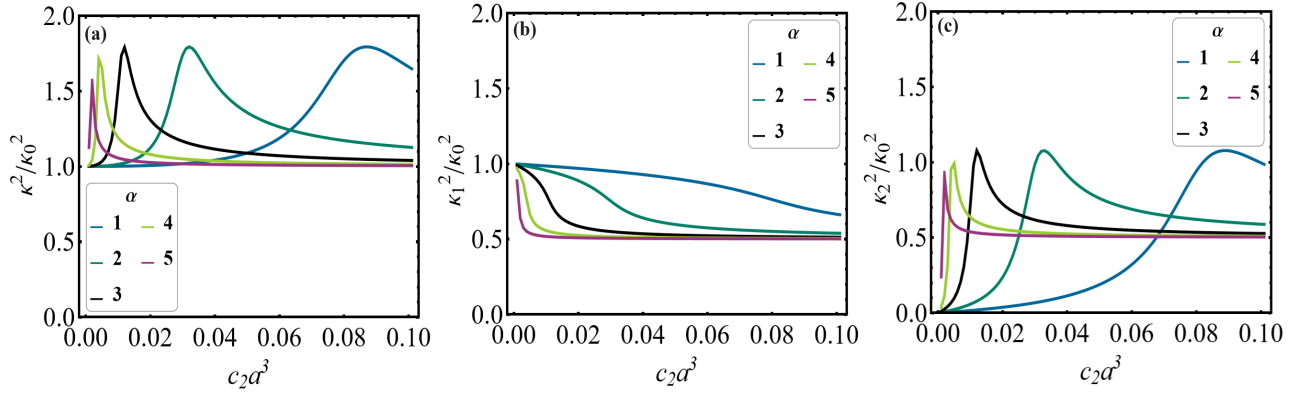

Figure 2: (Color online) Inverse square of the screening length, $\kappa^2$, divided by $\kappa_0^2$, Eq. 25, and its two components $\kappa_1^2$ and $\kappa_2^2$ as defined in the main text, as a function of $c_2$ for different values of adsorption energy $\alpha = 1, 2, 3, 4, 5$ and the Flory-Huggins interaction $\chi = -3$ at fixed $N_1 = 1$. The $\kappa_1^2$ term is basically the Debye length with CR charges, smaller than the bare charges so that $\kappa_1^2/\kappa_1^2 \leq 1$. The $\kappa_2^2$ term stems from the fluctuations in the CR charge and shows a "resonance" peak depending on $\alpha, \kappa$ values.

that case, the pure inverse square of the Debye length, $\kappa_0^2$, would be

$$\kappa_0^2(c_2)/(4\pi\ell_B) \equiv \kappa^2(c_2)|_{\phi=0}/(4\pi\ell_B) \;=\; c_2 + N_1^2 c_1(c_2) = c_2 \left(1 + N_1\right), \tag{24}$$

where the $c_1$ was obtained from the electroneutrality condition, Eq. 22. The rescaled inverse square of the screening length can then be cast as $\kappa^2/\kappa_0^2$ and has the property of $\kappa^2/\kappa_0^2 \longrightarrow 1$ when charge regulation is absent.

Other possibilities for the proper normalization of the screening length are also possible, but we specifically focus on the limit of no charge regulation as a baseline.

We will split the inverse square of the screening length into two separate contributions, $\kappa^2 = \kappa_1^2 + \kappa_2^2$, so that

$$\kappa^2/\kappa_0^2 \equiv \frac{1 + N_1(1 - \phi(c_2))}{1 + N_1} + \frac{1}{(1 + N_1)(1 - \phi(c_2))} \left( \frac{1}{\frac{1}{\phi(c_2)(1-\phi(c_2))} + \chi} \right), \tag{25}$$

where $\kappa_1^2$ pertains to the first term, originating in the Debye screening length with CR charges, and $\kappa_2^2$ to the second one, whose origin is the charge fluctuation due to the CR mechanism, Eq. 19.

There are several parameters characterizing the present CR model: the adsorption energy $\alpha$, the Flory-Huggins parameter $\chi$, and the number of CR sites on the macroions surface, $N_1$. In what follows we do not aim at the exhaustive numerical explorations but try to identify how the features of the screening are impacted by the variation in these parameters.

The $\kappa^2(c_2)$, composed of two terms, $\kappa_1^2$ and $\kappa_2^2$ in Eq. 25, is shown in Fig. 2. Clearly $\kappa^2(c_2)$ shows a "screening resonance" coinciding with the maximum in the CR charge fluctuations, $\kappa_2^2$, the second term in Eq. 25. This maximum persists even for small values of $\chi$, indicating that the "screening resonance" could be ubiquitous for CR systems. However, as we will see, the CR charge fluctuation term and the "screening resonance" scale inversely with the number of sites, $N_1$, and is thus bound to be quenched for macroions with many CR sites. The first term in Eq. 25, $\kappa_1^2$, is due to the intrinsic screening where the effective charges of the counterion and macroion are given by the CR mechanism, Eq. 19, and are a consequence of the density fluctuations of the ions. Since the effective charge in a CR system is always smaller than the maximal charge, the inverse square of the screening length is therefore smaller than the Debye screening length. This is clearly displayed in Fig. 2.

The salient features of CR screening, described above, vary with the CR parameters, $\alpha$ and $\chi$, see Fig. 3. The variation is predictable and amounts to the position of the "screening resonance" peak and its width. For small negative $\chi$ the peak is very broad and moves towards smaller values of $c_2$, as $\alpha$ becomes larger, becoming sharper at the same time. This remains true for $\chi$ getting more negative except that the "screening resonance" peak becomes sharper with more negative $\chi$. Depending on the model of CR at the macroion surface, *i.e.*, the value of $\chi$, the peak of the "screening resonance" can be quite high, implying an anomalously large screening that would prevent the electric field from entering the macroion solution. So even if the solution does not have a large concentration of the mobile ions that would screen the field, the CR process itself would provide the screening mechanism irrespective of the low concentration of the mobile charge carriers.

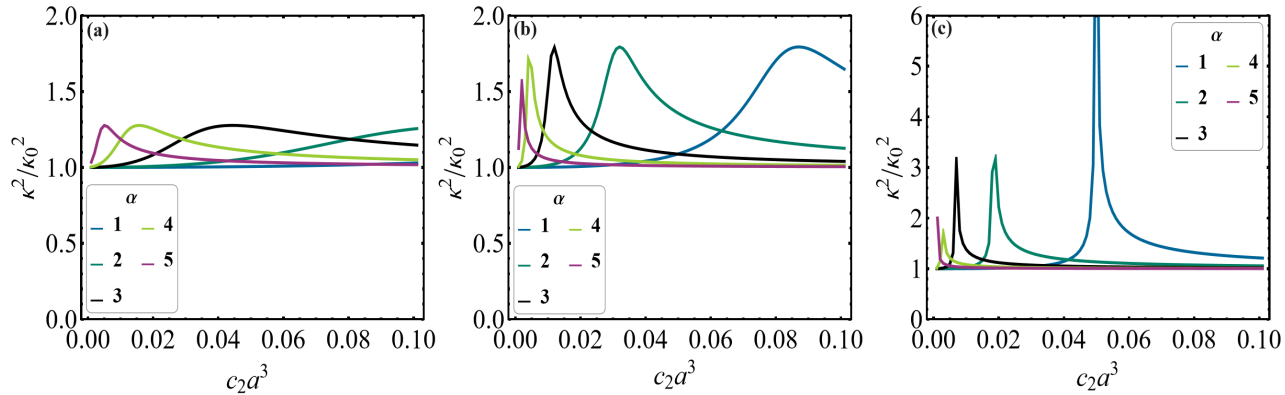

Figure 3: (Color online) Inverse square of the screening length, $\kappa^2$, divided by the standard Debye length, $\kappa_0^2$, Eq. 25, as a function of $c_2$ for different values of adsorption energy $\alpha = 1, 2, 3, 4, 5$ and the Flory-Huggins interaction $\chi = -2(a), -3(b), -4(c)$ at fixed $N_1 = 1$. The "screening resonance" peak gets sharper for larger negative $\chi$, while its position varies with $\alpha$. For this choice of $N_1$ the ratio $\kappa^2/\kappa_0^2 \leq 1$. For small negative $\chi$ the screening resonance can effectively prevent the electric field from penetrating the macroion solution.

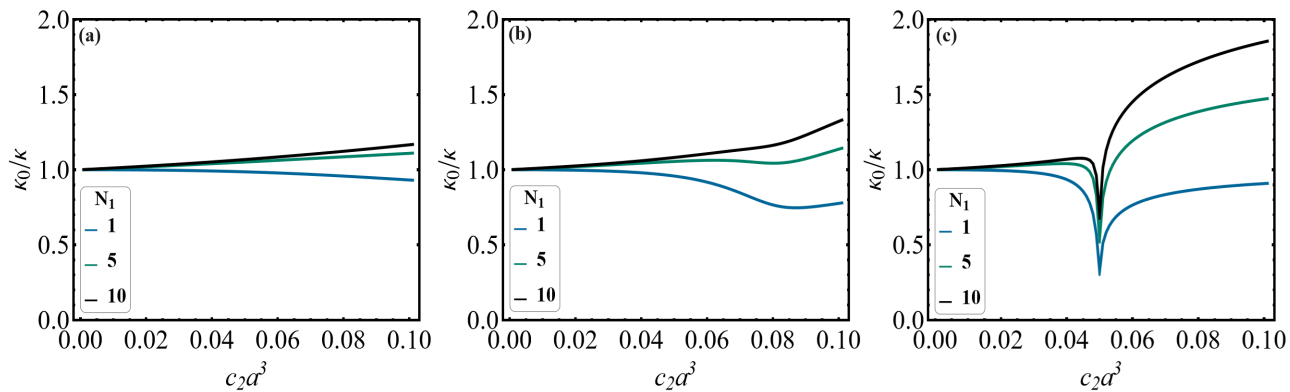

Figure 4: (Color online)(a) The ratio of Debye length and inverse screening length $(\kappa_0/\kappa)$ as a function of $c_2$ for different values of $N_1 = 1, 5, 10$ and the Flory-Huggins interaction $\chi = -2(a), -3(b), -4(c)$ at fixed adsorption energy $\alpha = 1$. For $N_1 = 1$ the screening length is smaller than the Debye length because of the contribution of the $\kappa_2$ term to the overall screening, Eq. 25. At larger $N_1 \gg 1$ the $\kappa_1$ term starts to dominate and the screening length becomes larger than the Debye length. This is the underscreening of a CR macroion solution which results from the smaller effective charge of the CR macroions and not from their smaller effective density.

The variation of the screening length with the number of CR sites, $N_1$, reveals the importance of various components of the screening phenomenon in CR solutions by plotting the ratio of the screening length, $\kappa^{-1}$, and Debye length, $\kappa_0^{-1}$, Fig. 4. Clearly, for $N_1 = 1$ the $\kappa^2/\kappa_0^2$ is always above one, so that the screening length is smaller than the Debye length indicating an *overscreening*. But this case is rather artificial since a macroion would usually display many dissociation sites, so necessarily $N_1 \geq 1$. As can be seen for $N_1 = 5, 10$ the screening length can be larger than the Debye length, indicating an *underscreening*. This underscreening stems from the first term in Eq. 25, while the contribution of the CR charge fluctuation term, the second term in Eq. 19, obviously fades away for $N_1 \gg 1$, as the first term asymptotes at $(1 - \phi(c_2))$ for $N_1 \gg 1$. It corresponds to pure "mean field" charge regulation, where the effective charges of the counterion and macroion are given by the mean-field CR equations. One could thus refer to this term as a "mean-field" screening term. The second term asymptotes to zero for $N_1 \gg 1$ and thus ceases to have an effect on the screening of a CR solution. The scaling of this term with $N_1$ can be understood intuitively from Eq. 19: the fluctuations of the charge, $\left(\langle \tilde{Q}_2^2 \rangle - \langle \tilde{Q}_2 \rangle^2\right)_{c_2}$, being proportional to the fluctuations of the number of cations at the CR surface of the macroion, scale inversely as the number of sites and thus become subdominant for a large number of sites.

While our quantitative analysis is based on the Fowler-Guggenheim-Frumkin isotherm, other models are of course possible [39, 40]. Qualitatively the details of the adsorption model do not matter as long as the model allows for a relatively sudden onset of the adsorption.

Finally, the "underscreening" in concentrated electrolytes that have been a vigorously pursued topic experimentally as well as theoretically [47–52] is probably not related to charge regulation, as it exists also in monovalent salts where the ionic charges are not charge regulated. The effects that we describe should thus be relevant in more complicated macromolecular solutions of charge-regulated macroions, such as proteins and/or nanoparticles. Recent advances in the simulation techniques that include different types of CR mechanisms that are being developed and/or are available [13, 53–60] should allow for a detailed comparison with the present theoretical results.

## V.  ACKNOWLEDGMENT

RP would like to thank Dr. Yael Avni for illuminating discussions on the charge regulation problem in general and its contribution to the screening properties in particular. RP acknowledges funding from the Key Project No. 12034019 of the National Natural Science Foundation of China.

[1] J. N. Israelachvili, Intermolecular and Surface Forces (Academic Press, Amsterdam, 2011).
[2] R. H. French, V. A. Parsegian, R. Podgornik, R. F. Rajter, A. Jagota, J. Luo, D. Asthagiri, M. K. Chaudhury, Y.-m. Chiang, S. Granick, S. Kalinin, M. Kardar, R. Kjellander, D. C. Langreth, J. Lewis, S. Lustig, D. Wesolowski, J. S. Wettlaufer, W.-Y. Ching, M. Finnis, F. Houlihan, O. A. von Lilienfeld, C. J. van Oss, and T. Zemb, Long range interactions in nanoscale science, Rev. Mod. Phys. **82**, 1887 (2010).
[3] M. Muthukumar, Physics of Charged Macromolecules: Synthetic and Biological Systems (Cambridge University Press, 2023).
[4] T. Markovich, D. Andelman, and R. Podgornik, Charged membranes: Poisson-boltzmann theory, dlvo paradigm and beyond, in Handbook of Lipid Membranes, edited by C. Safynia and J. Raedler (Taylor and Francis, 2018).
[5] S. Buyukdagli, Impact of the inner solute structure on the electrostatic mean-field and strong-coupling regimes of macromolecular interactions, Phys. Rev. E **107**, 064604 (2023).
[6] A. Naji, M. Kanduč, J. Forsman, and R. Podgornik, Perspective: Coulomb fluids—weak coupling, strong coupling, in between and beyond, J. Chem. Phys. **139**, 150901–13 (2013).
[7] S. Buyukdagli and R. Podgornik, Like-charge polymer-membrane complexation mediated by multivalent cations: One-loop-dressed strong coupling theory, J. Chem. Phys. **151**, 094902 (2019).
[8] R. Blossey, The Poisson-Boltzmann Equation: An Introduction, Springer Briefs in Physics (Springer Nature, Springer Nature Tiergartenstrasse 15 – 17 69121 Heidelberg Germany, 2023).
[9] E. J. Verwey and J. T. G. Overbeek, Theory of the stability of Lyophobic Colloids (Elsevier, Amsterdam, 1948).
[10] M. Borkovec, B. Jönsson, and G. J. M. Koper, Ionization processes and proton binding in polyprotic systems: Small molecules, proteins, interfaces, and polyelectrolytes, in Surface and Colloid Science, edited by E. Matijević (Springer, 2001) pp. 99–340.
[11] Y. Avni, D. Andelman, and R. Podgornik, Charge regulation with fixed and mobile charged macromolecules, Curr. Opin. Electrochem. **13**, 70–77 (2019).
[12] G. Trefalt, T. Palberg, and M. Borkovec, Curr. Opin. Colloid Interface Sci. **27**, 9 (2017).
[13] J. Yuan, K. Takae, and H. Tanaka, Impact of charge regulation on self-assembly of zwitterionic nanoparticles, Phys. Rev. Lett. **128**, 158001 (2022).
[14] H. Ruixuan, A. Majee, J. Dobnikar, and R. Podgornik, Electrostatic interactions between charge regulated spherical macroions (2023), arXiv:2308.03254 [cond-mat.soft].
[15] T. Obstbaum and U. Sivan, Thermodynamics of charge regulation near surface neutrality, Langmuir **38**, 8477 (2022).
[16] T. Obstbaum and U. Sivan, Charge regulation indicates water expulsion from silica surface by cesium cations, J. Colloid Interface Sci. **638**, 825 (2023).
[17] M. Lund and B. Jönsson, Charge regulation in biomolecular solution, Q. Rev. Biophys. **46**, 265 (2013).
[18] T. Pálmadóttir, A. Malmendal, T. Leiding, M. Lund, and S. Linse, Charge regulation during amyloid formation of $\alpha$-synuclein, Journal of the American Chemical Society **143**, 7777 (2021).
[19] M. Ghasemi and R. G. Larson, Role of electrostatic interactions in charge regulation of weakly dissociating polyacids, Prog. Polym. Sci. **112**, 101322 (2021).
[20] M. Lund, T. Akesson, and B. Jönsson, Enhanced protein adsorption due to charge regulation, Langmuir **21**, 8385 (2005).
[21] F. L. B. da Silva and B. Jönsson, Polyelectrolyte–protein complexation driven by charge regulation, Soft Matter **5**, 2862 (2009).

[22] A. Neamtu, F. Mocci, A. Laaksonen, and F. L. B. da Silva, Towards an optimal monoclonal antibody with higher binding affinity to the receptor-binding domain of sars-cov-2 spike proteins from different variants, Colloids Surf. B Biointerfaces **221**, 112986 (2023).

[23] G. S. Longo, M. Olvera de la Cruz, and I. Szleifer, Molecular theory of weak polyelectrolyte gels: The role of ph and salt concentration, Macromolecules **44**, 147 (2011).

[24] P. K. Jha, P. S. Desai, J. Li, and R. G. Larson, ph and salt effects on the associative phase separation of oppositely charged polyelectrolytes, Polymers **6**, 1414 (2014).

[25] A. Salehi and R. G. Larson, A molecular thermodynamic model of complexation in mixtures of oppositely charged polyelectrolytes with explicit account of charge association/dissociation, Macromolecules **49**, 9706–9719 (2016).

[26] R. J. Nap, B. Qiao, L. C. Palmer, S. I. Stupp, M. Olvera de la Cruz, and I. Szleifer, Acid-base equilibrium and dielectric environment regulate charge in supramolecular nanofibers, Frontiers in Chemistry **10**, 852164 (2022).

[27] F. L. B. da Silva, P. Derreumaux, and S. Pasquali, Protein-rna complexation driven by the charge regulation mechanism, Biochem. Biophys. Res. Commun. **498**, 264 (2018).

[28] G. L. Celora, R. Blossey, A. Münch, and B. Wagner, Counterion-controlled phase equilibria in a charge-regulated polymer solution (2023), arXiv:2307.03706 [cond-mat.soft].

[29] S. Kumari, S. Dwivedi, and R. Podgornik, On the nature of screening in voorn–overbeek type theories, J. Chem. Phys. **156**, 244901 (2022).

[30] Y. Groda, M. Dudka, G. Oshanin, A. A. Kornyshev, and S. Kondrat, Ionic liquids in conducting nanoslits: how important is the range of the screened electrostatic interactions?, J. Phys.: Condens. Matter **34**, 26LT01 (2022).

[31] Z. A. Goodwin, M. McEldrew, J. Pedro de Souza, M. Z. Bazant, and A. A. Kornyshev, Gelation, clustering, and crowding in the electrical double layer of ionic liquids, J. Chem. Phys. **157**, 094106 (2022).

[32] T. Markovich, D. Andelman, and R. Podgornik, Complex fluids with mobile charge-regulating macro-ions, EuroPhys. Lett. **120**, 26001 (2018).

[33] Y. Avni, T. Markovich, R. Podgornik, and D. Andelman, Charge regulating macro-ions in salt solutions: screening properties and electrostatic interactions, Soft Matter **14**, 6058 (2018).

[34] M. Lund and B. Jönsson, On the charge regulation of proteins, Biochemistry **44**, 5722 (2005).

[35] A. C. Maggs, Dynamics of a local algorithm for simulating coulomb interactions, J. Chem. Phys. **117**, 1975 (2002).

[36] A. C. Maggs and R. Podgornik, General theory of asymmetric steric interactions in electrostatic double layers, Soft Matter **12**, 1219–1229 (2016).

[37] M. Muthukumar, J. Hua, and A. Kundagrami, Charge regularization in phase separating polyelectrolyte solutions, J. Chem. Phys. **132**, 084901–9 (2010).

[38] D. Harries, R. Podgornik, V. A. Parsegian, E. Mar-Or, and D. Andelman, Ion induced lamellar-lamellar phase transition in charged surfactant systems, J. Chem. Phys. **124**, 224702 (2006).

[39] L. Koopal, W. Tan, and M. Avena, Equilibrium mono- and multicomponent adsorption models: From homogeneous ideal to heterogeneous non-ideal binding, Adv. Colloid Interface Sci. **280**, 102138 (2020).

[40] M. Mozaffari Majd, V. Kordzadeh-Kermani, V. Ghalandari, A. Askari, and M. Sillanpää, Adsorption isotherm models: A comprehensive and systematic review (2010-2020), Sci. Total Environ. **812**, 151334 (2022).

[41] I. Teraoka, *Polymer Solutions: An Introduction to Physical Properties* (Wiley Interscience, John Wiley and Sons Inc., New York, 2002).

[42] R. Podgornik, General theory of charge regulation and surface differential capacitance, J. Chem. Phys. **149**, 104701 (2018).

[43] R. Blossey and R. Podgornik, A comprehensive continuum theory of structured liquids, Journal of Physics A: Mathematical and Theoretical **56**, 025002 (2023).

[44] T. L. Hill, *Statistical mechanics: Principles and Selected Applications*, Dover Books on Physics (Dover Publications, Mineola, NY, 1987).

[45] R. Zia, E. F. Redish, and S. R. McKay, Making sense of the legendre transform, Am. J. Phys. **77**, 614–622 (2009).

[46] M. Han, H. Kim, C. Leal, M. Negrito, J. D. Batteas, and R. M. Espinosa-Marzal, Insight into the electrical double layer of ionic liquids revealed through its temporal evolution, Adv. Mater. Interfaces **7**, 2001313 (2020).

[47] A. A. Lee, C. S. Perez-Martinez, A. M. Smith, and S. Perkin, Underscreening in concentrated electrolytes, Faraday Discuss. **199**, 239 (2017).

[48] Z. A. H. Goodwin and A. A. Kornyshev, Underscreening, overscreening and double-layer capacitance, Electrochem. Commun. **82**, 129 (2017).

[49] B. Rotenberg, O. Bernard, and J.-P. Hansen, Underscreening in ionic liquids: a first principles analysis, J. Phys.: Condens. Matter **30**, 054005 (2018).

[50] R. M. Adar, S. A. Safran, H. Diamant, and D. Andelman, Screening length for finite-size ions in concentrated electrolytes, Phys. Rev. E **100**, 042615 (2019).

[51] M. McEldrew, Z. A. H. Goodwin, S. Bi, M. Z. Bazant, and A. A. Kornyshev, Theory of ion aggregation and gelation in super-concentrated electrolytes, J. Chem. Phys. **152**, 234506 (2020).

[52] S. A. Safran and P. A. Pincus, Scaling perspectives of underscreening in concentrated electrolyte solutions, Soft Matter **19**, 7907 (2023).

[53] B. K. Radak, C. Chipot, D. Suh, S. Jo, W. Jiang, J. C. Phillips, K. Schulten, and B. Roux, Constant-ph molecular dynamics simulations for large biomolecular systems, Journal of Chemical Theory and Computation **13**, 5933 (2017).

[54] J. Landsgesell, P. Hebbeker, O. Rud, R. Lunkad, P. Košovan, and C. Holm, Grand-reaction method for simulations of ionization equilibria coupled to ion partitioning, Macromolecules **53**, 3007 (2020).
[55] J. Landsgesell, L. Nová, O. Rud, F. Uhlík, D. Sean, P. Hebbeker, C. Holm, and P. Košovan, Simulations of ionization equilibria in weak polyelectrolyte solutions and gels, Soft Matter **15**, 1155 (2019).
[56] T. Curk, J. Yuan, and E. Luijten, Accelerated simulation method for charge regulation effects, J. Chem. Phys. **156**, 044122 (2022).
[57] N. Aho, P. Buslaev, A. Jansen, P. Bauer, G. Groenhof, and B. Hess, Scalable constant ph molecular dynamics in GROMACS, J. Chem. Theory Comput. **18**, 6148 (2022).
[58] A. Bakhshandeh, D. Frydel, and Y. Levin, Theory of charge regulation of colloidal particles in electrolyte solutions, Langmuir **38**, 13963 (2022).
[59] P. Buslaev, N. Aho, A. Jansen, P. Bauer, B. Hess, and G. Groenhof, Best practices in constant ph MD simulations: Accuracy and sampling, J. Chem. Theory Comput. **18**, 6134 (2022).
[60] A. Bakhshandeh, A. P. Dos Santos, and Y. Levin, Interaction between charge-regulated metal nanoparticles in an electrolyte solution, J. Phys. Chem. B **124**, 11762 (2020).